\documentclass[prb,twocolumn,amssymb, nobibnotes, aps, superscriptaddress]{revtex4-2}
\usepackage{amsmath}
\usepackage{graphicx}
\usepackage{bm}
\usepackage{graphicx}
\usepackage{float}
\usepackage{subfigure}
\usepackage{makecell}
\usepackage{booktabs}
\usepackage{natbib}
\usepackage{color}
\usepackage[colorlinks=true, allcolors=blue]{hyperref}
\usepackage{wasysym}
\usepackage[utf8]{inputenc}

\usepackage{appendix}
\usepackage{changes}

\begin{document}
\affiliation{Key Laboratory of Artificial Micro- and Nano-structures of Ministry of Education and School of Physics and Technology, Wuhan University, Wuhan 430072, China}
\affiliation{Lanzhou Center of Theoretical Physics $\&$ Key Laboratory for Quantum Theory and Applications of the Ministry of Education $\&$ Key Laboratory of Theoretical Physics of Gansu Province $\&$ School of Physical Science and Technology, Lanzhou University, Lanzhou 730000, China}
\affiliation{AI for Science Institute, Beijing, 100080, China}
\affiliation{Quantum Science Center of Guangdong-Hongkong-Macao Greater Bay Area (Guangdong), Shenzhen 518045, China}
\title{Electronic states and quantum transport in bilayer graphene Sierpinski-carpet fractals}%

\author{Xiaotian Yang}
\affiliation{Key Laboratory of Artificial Micro- and Nano-structures of Ministry of Education and School of Physics and Technology, Wuhan University, Wuhan 430072, China}
\author{Weiqing Zhou}
\affiliation{AI for Science Institute, Beijing, 100080, China}
\author{Qi Yao}
\affiliation{Quantum Science Center of Guangdong-Hongkong-Macao Greater Bay Area (Guangdong), Shenzhen 518045, China}
\author{Yunhai Li}
\affiliation{Key Laboratory of Artificial Micro- and Nano-structures of Ministry of Education and School of Physics and Technology, Wuhan University, Wuhan 430072, China}
\author{Yunhua Wang}
\email[Corresponding author: ]{wangyunhua@lzu.edu.cn}
\affiliation{Lanzhou Center of Theoretical Physics $\&$ Key Laboratory for Quantum Theory and Applications of the Ministry of Education $\&$ Key Laboratory of Theoretical Physics of Gansu Province $\&$ School of Physical Science and Technology, Lanzhou University, Lanzhou 730000, China}
\affiliation{Key Laboratory of Artificial Micro- and Nano-structures of Ministry of Education and School of Physics and Technology, Wuhan University, Wuhan 430072, China}
\author{Shengjun Yuan}%
\email[Corresponding author: ]{s.yuan@whu.edu.cn}
\affiliation{Key Laboratory of Artificial Micro- and Nano-structures of Ministry of Education and School of Physics and Technology, Wuhan University, Wuhan 430072, China}
\affiliation{Wuhan Institute of Quantum Technology, Wuhan 430206, China}

\begin{abstract}
We construct Sierpinski-carpet (SC) based on AA or AB bilayer graphene by atom vacancies, namely, SC-AA and SC-AB, to investigate the effects of interlayer coupling on the electronic properties of fractals. Compared with monolayer graphene SC, their density of states have similar features, such as Van-Hove singularities and edge states corresponding to the central peaks near zero energy, but remarkable energy broadening of edge states emerges in SC-AA(AB). Calculated conductance spectrum shows that the conductance fluctuations still hold the Hausdorff fractal dimension behavior even with the interlayer coupling. Thus, the high correlation between quantum conductance and fractal geometry dimension is not affected by the interlayer coupling in bilayer graphene SC. We further reveal the quasi-eigenstates in fractal-like pressure-modulated bilayer graphene, namely, SC-pAA and SC-pAB. Numerical results show that the density of states of SC-pAA(pAB) show an asymptotic behavior to those of SC-AA(AB) especially for high energy quasi-eigenstates. Within a certain energy range, stronger pressure can lead to stronger localization, forming an efficient fractal space.
\end{abstract}
\maketitle

\section{INTRODUCTION}\label{sec:Introduction}
Fractal is a unique structure for its fascinating self-similarity and non-integer Hausdorff dimension $d_{\rm H}$ \cite{gefen1984phase,geometryoffractal,pietronero2012fractals,feder2013fractals}. The two intrinsic characteristics in this unconventional system enable exotic and interesting physical features on electronic energy spectrum statistics \cite{yao2023energy,iliasov2020linearized,iliasov2019power,kosior2017localization,hernando2015spectral}, quantum transport \cite{2016transport,yang2020confined,yang2022electronic,iliasov2020hall,fremling2020existence,bouzerar2020quantum,han2019universal},
plasmons \cite{westerhout2018plasmon}, flat bands \cite{nandy2021controlled,nandy2015flat,nandy2015engineering,pal2018flat}, topological phases \cite{sarangi2021effect,fischer2021robustness,PhysRevResearch.2.023401,pai2019topological,brzezinska2018topology}, enhanced superconductivity \cite{iliasov2023strong} and modified super-area law of entanglement entropy \cite{zhou2023entanglement}. Experimentally, nanoscale fractal, such us Sierpinski carpet (SC) and Sierpinski gasket are mainly created by the bottom-up nanofabrication methods, including molecular self-assembly \cite{jiang2017constructing,nieckarz2016chiral,zhang2015controlling,sun2015surface,tait2015self,shang2015assembling,newkome2006nanoassembly}, chemical reactions \cite{zhang2016robust}, template packings \cite{li2017packing} and atomic manipulations in a scanning tunneling microscope \cite{PhysRevLett.126.176102,jiang2021direct,kempkes2019design}. Recently, SC photonic lattices are also created to investigate the photon evolution \cite{xu2021shining}. Top-down external field modulation is another feasible method for generating large-scale fractal structures. Especially, graphene lattice is much more easier to form an effective fractional dimension with a small electric field, compared with square lattice materials \cite{yang2020confined}. In addition, there are some unique physical properties in graphene fractal systems. For instance, the geometry dimension of monolayer graphene SC is characterized by quantum conductance fluctuations \cite{2016transport,yang2020confined}. Besides, the eigenstates of graphene fractals exhibit various localized distributions in real space, and the edge states induced by zigzag terminations are localized at the hole boundaries of the graphene SC, forming a special states distribution \cite{yang2020confined}. In functionalized graphene SC, there are two special energy windows, where holes are mainly located inside functionalized region and electrons are mainly located inside fractal region \cite{yang2022electronic}.

Different from the electronic structure of monolayer graphene, bilayer graphene with AA or AB stacking has a parabolic dispersion at low energy range owing to the interlayer coupling. Recent theoretical calculations and experimental researches show that external vertical pressure can enhance the interlayer coupling and change the physical properties of few-layer graphene, like electronic structure \cite{ge2021emerging,chittari2018pressure,yankowitz2016pressure,proctor2009high,yankowitz2018dynamic}, Raman spectrum \cite{proctor2009high, yankowitz2018dynamic,nicolle2011pressure, machon2018raman}, magnetism \cite{castro2011effect,chen2022first, pant2022phase}, phase transition \cite{pant2022phase,sanchez2023pressure,padhi2019pressure}, and superconductivity \cite{yankowitz2019tuning}. Especially, it has been proved that the interlayer interactions in certain regions can be controlled by locally modifying the interlayer separation by applying a pressure from a scanning tunneling microscopy tip \cite{yankowitz2016pressure}. From an natural extension of graphene fractals, we want to know how the electronic states and quantum transports are in bilayer graphene Sierpinski-carpet fractals formed by atom vacancies, namely, SC-AA(AB). Owing to the high tunability of the interlayer coupling by pressure, we further want to ask, how the electronic states are in fractal-like pressure modulation AA and AB bilayer graphene (namely, SC-pAA and SC-pAB, respectively), and what the difference is among the four types of bilayer graphene fractals.

\begin{figure*}[tbp]
\centering
\includegraphics[width=14cm]{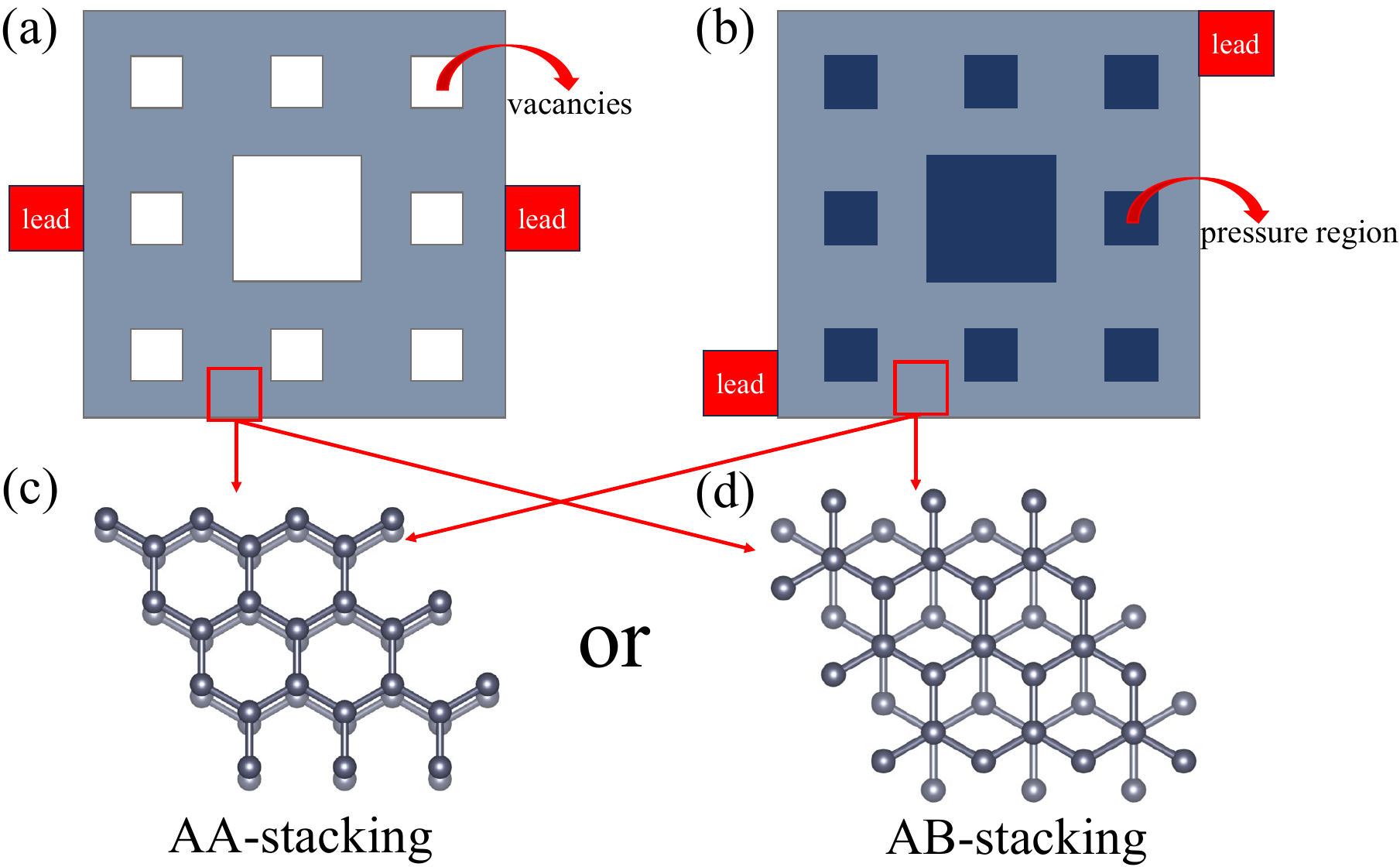}
    \caption{(a) Schematic diagram of AA or AB stacking graphene SC sample generated by atom vacancies with the iteration number $I =2$ and the square width $W=32.5a$ with $a$ as the lattice constant of graphene. (b) Schematic diagram of AA or AB stacking graphene SC sample generated by local pressure modulations in the dark regions named as Area II. the other light region is named as Area I. The atomic structures of AA and AB stacking bilayer graphene are shown in (c) and (d), respectively. Different position configurations of leads, called central and diagonal leads, are applied to the sample for calculations on quantum conductance.}
    \label{sample}
\end{figure*}

In this work, we investigated the electronic states and quantum conductance fluctuations in SC-AA(AB). Remarkable energy broadening of quasi-eigenstates around zero energy as edge states are observed in the calculated results of density of states (DOS) and real-space distributions of probability density. The quantum conductance fluctuations hold the Hausdorff fractal dimension behavior similar to that of monolayer graphene fractals, even in the presence of the interlayer coupling in bilayer graphene fractals. We also reveal the electronic states in pressure-modulated bilayer graphene fractals including SC-pAA and SC-pAB structures. The DOS results of SC-pAA(pAB) especially for high energy quasi-eigenstates show an asymptotic behavior to those of SC-AA(AB) as pressure increases. Our analyses on real-space distributions of normalized probability density also verify the DOS results. Besides, within a certain pressure range, stronger pressure can lead to stronger localization, forming a more efficient fractal space in SC-pAA(pAB). However, even for these high energy quasi-eigenstates, their DOS in SC-pAA(pAB) can not exactly replicate the same spectrum of SC-AA(AB) within the experimental pressure range.

The paper is organized as follows. In Sec.~\ref{sec:Model}, we describe the tight-binding model and details of applied numerical methods. In Sec.~\ref{sec:Results}, we perform the calculations and show the results of electronic properties and quantum transport for SC-AA(AB) and SC-pAA(pAB), including density of states, quasi-eigenstates, quantum conductance, and box-counting analysis of conductance fluctuation. A brief summary is given in Sec.~\ref{sec:Summary}.

\section{MODEL AND METHODS}\label{sec:Model}
We investigate two types of bilayer graphene SC structures: (i) the first type is generated by atom vacancies, as shown in Fig.~\ref{sample}(a), and (ii) the second type is formed by external pressure modulation, as depicted in Fig.~\ref{sample}(b). For convenience, we label the AA(AB)-stacking graphene SC formed by atom vacancies as SC-AA(AB), and the AA(AB)-stacking graphene SC with pressure as SC-pAA(pAB). The structural parameters in Fig.~\ref{sample} is listed here. $I = 2$ is the iteration number, and $W = 32.5a$ is the sample width with $a=2.46$ ${\rm \AA}$ as the lattice constant of graphene. When SC changes from the $I$th iteration to the $(I+1)$-th iteration, the unit is replicated with ${\cal N}=8$ times larger in area and ${\cal L}=3$ times larger in width. The Hausdorff dimension is defined by $d_{\rm H}  \equiv \log_{{\cal L}}{\cal N}$ $\simeq 1.89$. 

The behaviors of electrons and holes in these SC systems are governed by the following tight-binding Hamiltonian
\begin{equation}
\begin{aligned}
H = -\sum_{\alpha}\sum_{i,j}t_{\alpha,ij}c^{\dag}_{i}c_{j}+\sum_{i}\varepsilon_{i}c^{\dag}_{i}c_{i},
\end{aligned}
\label{equation1}
\end{equation}
where $\varepsilon_{i}$ is the on-site energy at the $i$-th site. $c_i^{\dag}$ and $c_j$ are creation and annihilation operators. $t_{ij}$ is electron hopping between $i$-th and $j$-th sites, including intralayer ($\alpha=0$) and interlayer ($\alpha=1$) hopping. This tight-binding Hamiltonian is obtained from the maximally localized Wannier orbitals \cite{marzari2012maximally,fang2016electronic}. The intralayer hopping energy up to the third nearest neighbors for the graphene monolayer has the values of -2.8922, 0.2425, and -0.2656 eV, respectively. The interlayer hopping is a function of both distance and  orientation \cite{fang2016electronic}
\begin{equation}
\begin{aligned}
t_{\alpha=1,ij}(\boldsymbol{r})=&V_{0}(r)+V_{3}(r)[\cos(3\theta_{12}+\cos(3\theta_{21})]\\&+V_{6}(r)[\cos(6\theta_{12})+\cos(6\theta_{21})],
\end{aligned}
\label{equation2}
\end{equation}
where the three terms originate from the different angular momenta of the wave functions. $\boldsymbol{r}$ is the projection vector connecting two sites in different layers. $r=|\boldsymbol{r}|$ denote the projected distances. $\theta_{12}$ and $\theta_{21}$ are the angles between the projected interlayer bonds and the in-plane nearest neighbor bonds. The three radial functions involving ten hopping parameters with $\Bar{r}=r/a$ are given by 
\begin{equation}
\begin{aligned}
&V_{0}(r)=\lambda_{0}e^{-\xi_{0}\Bar{r}^{2}}\cos{(\kappa_{0}\Bar{r})},\\
&V_{3}(r)=\lambda_{3}\Bar{r}^{2}e^{-\xi_{3}(\Bar{r}-x_{3})^{2}},\\
&V_{6}(r)=\lambda_{6}e^{-\xi_{6}(\Bar{r}-x_{6})^{2}}\sin{(\kappa_{6}\Bar{r})}.
\end{aligned}
\label{equation3}
\end{equation}
For bilayer graphene, the interlayer compression is related to the external pressure $P$ according to the Murnaghan equation of state \cite{chittari2018pressure}
\begin{equation}
\begin{aligned}
P=A(e^{-B\eta}-1),
\end{aligned}
\label{equation4}
\end{equation}
where $\eta$ is defined by $\eta=1-(h/h_{0})$, $h$ and $h_{0}$ are the distance at finite and zero external pressure. $A=5.73$ GPa and $B=9.54$ are obtained through the DFT calculations. The vertical compression of bilayer graphene has a weak effect on the intralayer hoppings but significantly enhances interlayer coupling \cite{carr2018pressure}. The ten hopping parameters involving interlayer coupling in Eq.~\eqref{equation3} are obtained through the following quadratic fitting,
\begin{equation}
\begin{aligned}
y_{i}(\eta)=f_{i}^{(0)}-f_{i}^{(1)}\eta+f_{i}^{(2)}\eta^{2},
\end{aligned}
\label{equation5}
\end{equation}
where $y_{i}(i=1,...,10)$ denotes arbitrary one of the ten hopping parameters with the coefficients $f_{i}^{(0)}$, $f_{i}^{(1)}$, and $f_{i}^{(2)}$ listed in Table.~\ref{table1} \cite{carr2018pressure}.

\begin{table}[H]
    \centering
    \caption{The ten hopping parameters in the interlayer coupling model. All these values are in units of eV and take the form in Eq.~\eqref{equation5}.}
    \renewcommand{\arraystretch}{1.2}
    \setlength{\tabcolsep}{6mm}{
    \begin{tabular}{c r r r}
    \hline\hline
    \specialrule{0em}{1pt}{1pt}
    $i(y_{i})$& $f_{i}^{(0)}$ & $f_{i}^{(1)}$&$f_{i}^{(2)}$\\
    \specialrule{0em}{1pt}{1pt}
    \hline
    \specialrule{0em}{1pt}{1pt}
    $1(\lambda_{0})$&$0.310$&$-1.882$&$7.741$\\
    $2(\xi_{0})$&$1.750$&$-0.618$&$1.848$\\
    $3(\kappa_{0})$&$1.990$&$1.007$&$2.427$\\
    $4(\lambda_{3})$&$-0.068$&$0.399$&$-1.739$\\
    $5(\xi_{3})$&$3.286$&$-0.914$&$12.011$\\
    $6(x_{3})$&$0.500$&$0.322$&$0.908$\\
    $7(\lambda_{6})$&$-0.008$&$0.046$&$-0.183$\\
    $8(\xi_{6})$&$2.272$&$-0.721$&$-4.414$\\
    $9(x_{6})$&$1.217$&$0.027$&$-0.658$\\
    $10(\kappa_{6})$&$1.562$&$-0.371$&$-0.134$\\
    \specialrule{0em}{1pt}{1pt}
    \hline\hline
    \end{tabular}}
    \label{table1}
\end{table}

\begin{figure*}[tbp]
\centering
\includegraphics[width=1.0\textwidth]{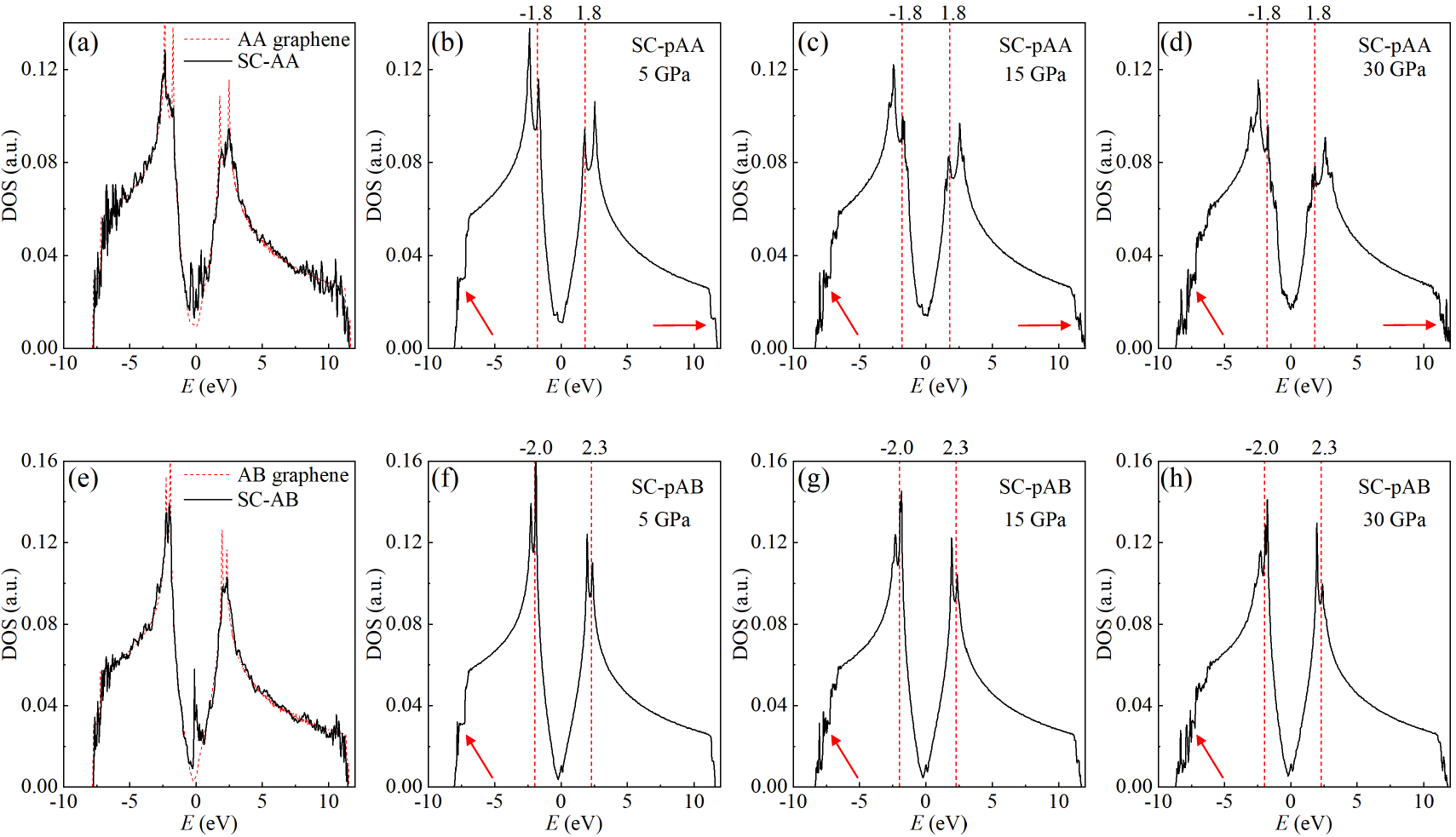}
\caption{DOS of SC-AA(AB) and pristine AA(AB)-stacking graphene are shown in (a) and (e). DOS of SC-pAA(pAB) for varying degrees of external pressure are shown in (b, f) 5 GPa, (c, g) 15 GPa, and (d, h) 30 GPa. The sample parameters are set as $W=297.5$a, and $I=4$.} 
\label{DOS}
\end{figure*}

Since numerical calculations based on exact diagonalization can only treat systems with site number less than ten thousand, we use tight-binding propagation method (TBPM) to calculate the electronic properties for large system with millions of sites, including density of states and quasi-eigenstates \cite{TBPM,li2023tbplas,zhou2023time}. We start the evolution of a quantum system with a random initial state $|\varphi(0)\rangle$, which is normalized superposition of all basis states $\sum_{n}A_{n}|n\rangle$. The DOS is calculated via Fourier transform of the correlation function \cite{TBPM,hams2000fast}:
\begin{equation}
\begin{aligned}
D(E)=\frac{1}{2\pi}\int_{-\infty}^{\infty}e^{iEt}\langle\varphi(0)|e^{-iHt}|\varphi(0)\rangle dt.
\end{aligned}
\label{equation6}
\end{equation}
After the Fourier transform of states at different time during the evolution $|\varphi(t)\rangle = e^{-iHt}|\varphi(0)\rangle$, we obtain the quasi-eigenstates $|\psi(E)\rangle$ by \cite{TBPM,kosloff1983fourier}
\begin{equation}
\begin{aligned}
|\psi(E)\rangle&=\frac{1}{2\pi}\int_{-\infty}^{\infty}dte^{iEt}|\varphi(t)\rangle\\&=\frac{1}{2\pi}\sum_{n}A_{n}\int_{-\infty}^{\infty}dte^{i(E-E_{n})t}|n\rangle\\&=\sum_{n}A_{n}\delta(E-E_{n})|n\rangle,
\end{aligned}
\label{equation7}
\end{equation}
which can be further normalized as
\begin{equation}
\begin{aligned}
|\psi(E)\rangle=\frac{1}{\sqrt{\sum_{n}|A_{n}|^{2}\delta(E-E_{n})}}\sum_{n}A_{n}\delta(E-E_{n})|n\rangle.
\end{aligned}
\label{equation8}
\end{equation}
For the finite fractal structure, one can make an average by different realizations of random coefficients $A_{n}$ to obtain more accurate results of $D(E)$ and $|\psi(E)\rangle$.

For the transport properties, we adopt the quantum transport simulator Kwant to do the numerical calculations \cite{groth2014kwant}. In Kwant, the system considered is treated as a scattering region. The scattering matrix $S_{ij}$ and the wave function inside the scattering region $\phi_{n}^{S}$ are the main raw output. They are calculated by matching the wave function in the lead to the wave function in the scattering region. After $S_{ij}$ is obtained, the quantum conductance $G_{ab}=dI_{a}/dV_{b}$ can be calculated by the Landauer formula
\begin{equation}
\begin{aligned}
G_{ab}=\frac{e^2}{h}\sum_{i\in{a},j\in{b}}|S_{ij}|^2,
\end{aligned}
\label{equation9}
\end{equation}
where $a$ and $b$ refer two electrodes.

\section{RESULTS AND DISCUSSION}\label{sec:Results}
\subsection{Density of states}
The fourth iteration $I=4$ for monolayer graphene SC is enough for the convergence of DOS calculations\cite{yang2020confined,yang2022electronic}. The number of sites ($\sim$ 410048 sites) for SC-AA and SC-AB is twice of that for monolayer graphene SC, and hence we only need average a small number of initial states within the TPBM method to explore electronic and transport characteristics in  SC-AA and SC-AB for $I=4$. The DOS of the SC-AA as a function of energy are shown in Fig.~\ref{DOS}(a). We can see that, due to the existence of the second and third nearest-neighbor hopping, the electron and hole states are not symmetrical in the energy spectrum. Two distinct Van Hove singularities appear around the $E=\pm2.9$ eV, similar to those of SC based on monolayer graphene. In fact, the DOS spectrum of pristine AA-stacking graphene exhibits four Van Hove singularities while the formation of fractal geometry caused by atomic vacancies leads to the merging of every two adjacent Van Hove singularities into a single peak, as shown in Fig.~\ref{DOS}(a). Besides, the central peak caused by edge states emerges due to the open boundaries. However, compared with the monolayer graphene SC \cite{yang2022electronic}, SC-AA exhibits an increased number of central peaks with an obvious broadening. In Fig.~\ref{DOS}(e), similar behavior is found in the DOS for SC-AB. The energy width of the central peak in SC-AB is also significantly enlarged. Here, we have used the Fermi energy as the zero energy reference point in the calculation.

\begin{figure*}[tbp]
\centering
\includegraphics[width=1.0\textwidth]{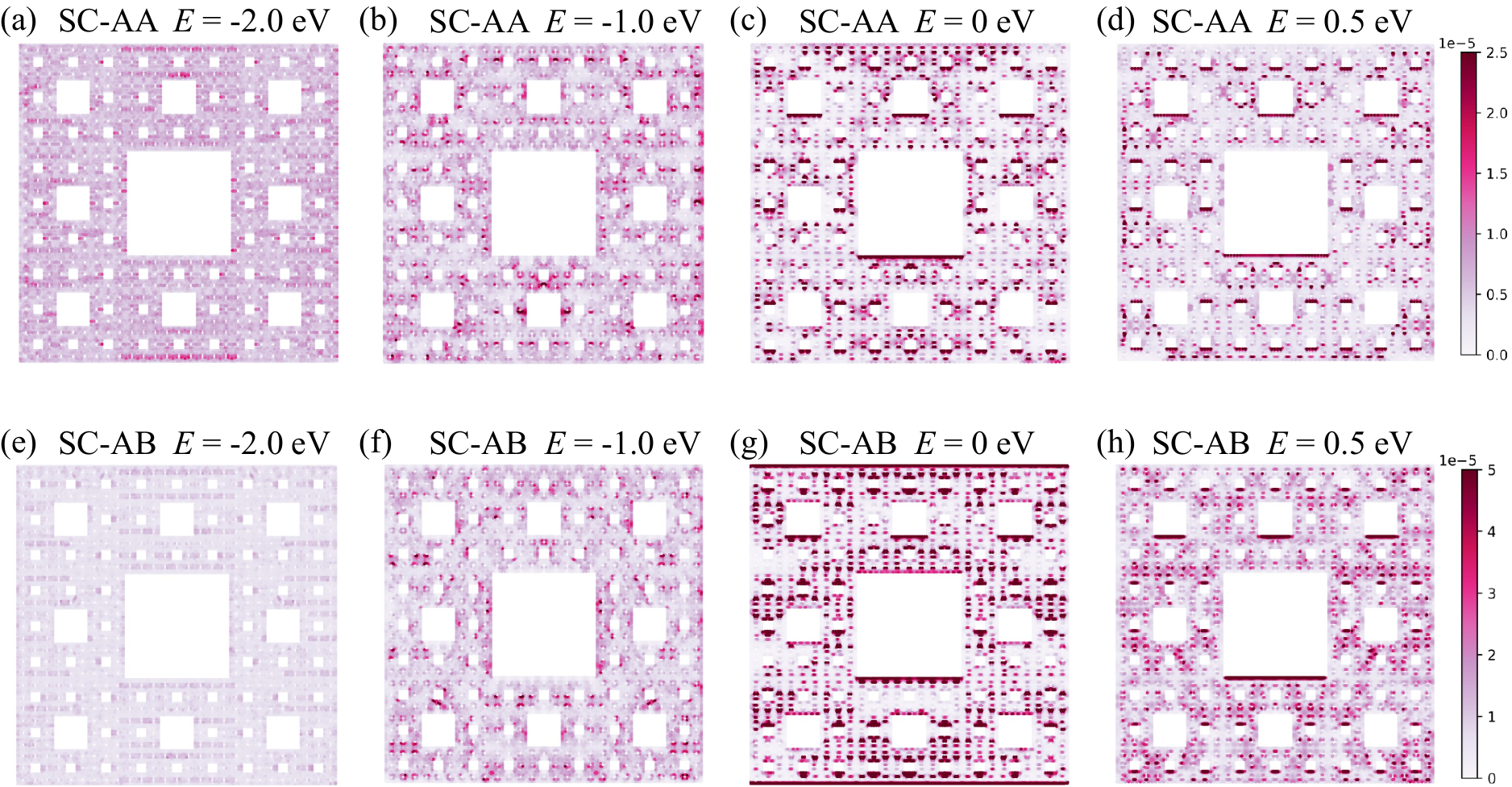}
\caption{The real-space distribution of quasi-eigenstates for SC-AA and SC-AB at (a, e) $E=-2$ eV, (b, f) $E=-1$ eV, (c, g) $E=0$ eV, and (d, h) $E=0.5$ eV. The sample parameters are set as $W=297.5$a, and $I=4$.} 
\label{eigen_sc-AA(AB)}
\end{figure*}

In order to investigate the behavior of electrons on fractal-like pressure modulated bilayer graphene, i.e., with Area II under pressure. We change the external pressure and examine the change of the DOS. For SC-pAA under 5 GPa pressure, the DOS is nearly identical to that of pristine AA-stacking bilayer graphene, with its four van Hove singularities (see Fig.~\ref{DOS}(b)). However, as the pressure increases, the two van Hove points gradually get close, and some small peaks appears around zero energy. In the high-energy region, the DOS becomes more chaotic like SC-AA, as shown by the red arrow in Figs.~\ref{DOS}(c) and~\ref{DOS}(d). Therefore, as the applied pressure increases, the pressured Area II acts as an insulating region preventing electrons and holes hopping from Area I to Area II. Naively, as the pressure grows infinitely larger, Area II will gradually be isolated from the whole system, and it will act like vacancies of SC-AA so that the DOS will be not smooth. Here, the maximum pressure applied during the DOS calculation is set to 30 GPa because there is no significant atomic restructuring of the graphene bilayer under this pressure value \cite{carr2018pressure}. In experiments, the diamond-anvil cell can apply a high pressure of up to 50 GPa to the suspended graphene bilayer \cite{tao2020raman}. As shown in Figs.~\ref{DOS}(f)-~\ref{DOS}(h), with increasing pressure, the DOS of SC-pAB also undergoes similar changes. There is a small peak at the zero energy point of the energy spectrum, and the DOS gradually becomes chaotic in the high-energy region. However, compared with SC-pAA, SC-pAB shows relatively lower sensitivity to pressure modulation. Based on these results we expect that the electronic states of SC-pAA(AB) will exhibit a fractal geometric distribution in some energy ranges, as will be confirmed in following Quasi-eigenstates (see Sec.~\ref{Qe}). However, within the considered pressure range, it cannot exactly replicate the same spectrum of SC-AA(AB), because Area II can be considered vacancies only if the pressure is infinite. In particular, the rate of DOS change in SC-pAB with pressure modulation is significantly slower than SC-pAA.

\begin{figure*}[tbp]
\centering
\includegraphics[width=0.98\textwidth]{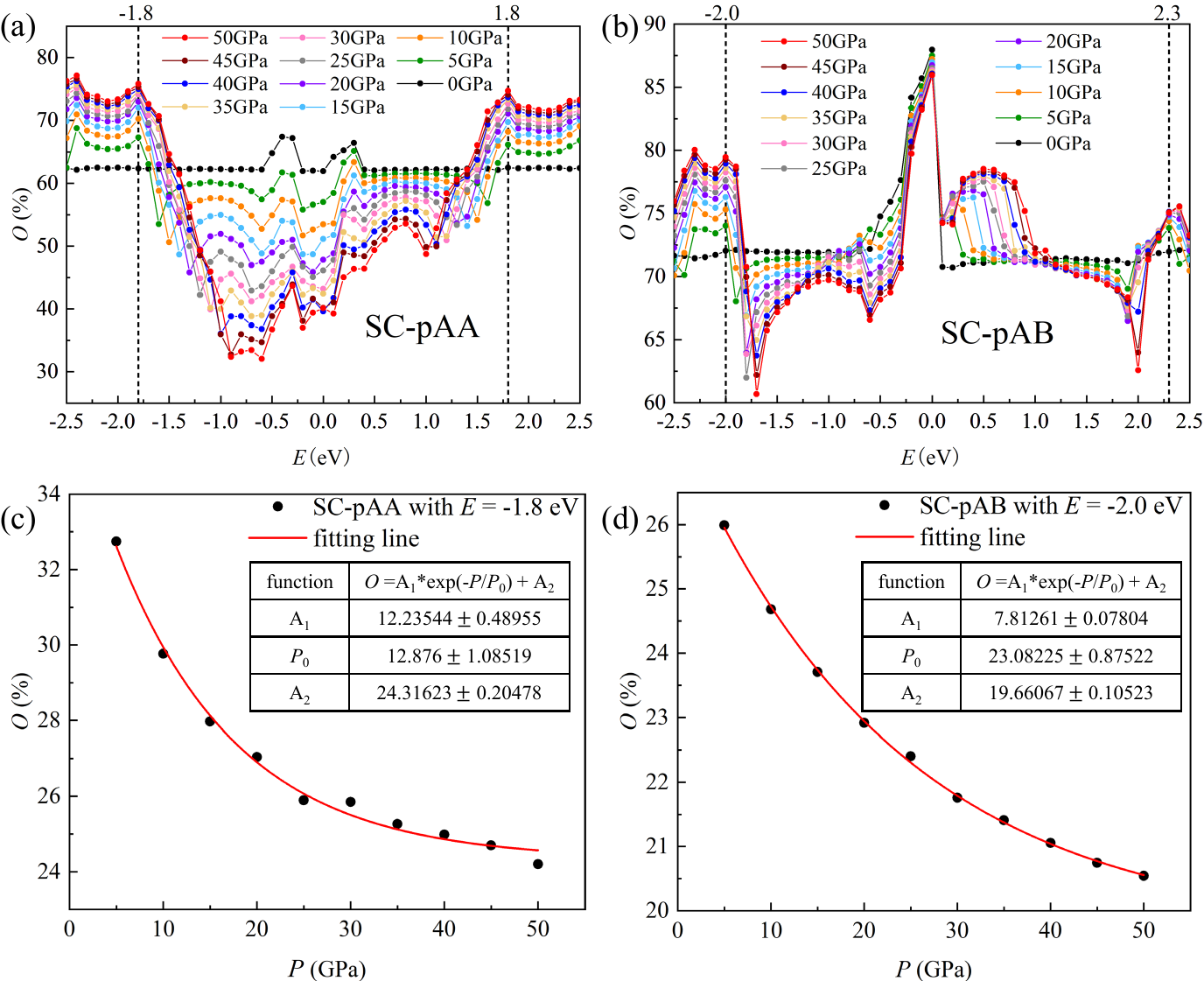}
\caption{(a) and (b) The occupation percentage $O$ of Area I in SC-pAA and SC-pAB under the different pressures and energies. (c) and (d) The occupation percentage $O$ of Area II in SC-pAA and SC-pAB under the different pressures at single energy point ($-1.8$ eV and $-2.0$ eV). The sample parameters are set as $W=297.5$a, and $I=4$.} 
\label{quasi_O}
\end{figure*}

\subsection{Quasi-eigenstates} \label{Qe}
Using Eq.~\eqref{equation7}, we calculate the quasi-eigenstates of SC-AA(AB) and plot the real-space distribution of their probability density in Fig.~\ref{eigen_sc-AA(AB)}. We first discuss the probability density distributions of high energy states in SC-AA, for example, the energy states at $E=-2$ eV and $E=-1$ eV in Figs.~\ref{eigen_sc-AA(AB)}(a) and~\ref{eigen_sc-AA(AB)}(b). Their nonzero probability densities are inside the fractal space (Area I). Some electronic states at $E=-2$ eV exhibit localization, and many electronic states for $E=-1$ eV are localized at the edge of holes formed by atomic vacancies. In fact, due to the existence of  atomic vacancies, electrons can only be confined in Area I. For the zero-energy states in Fig.~\ref{eigen_sc-AA(AB)}(c), these nonzero probability densities are obviously located at these zigzag terminations. In addition, these edge state still exist even for energy at $E=0.5$ eV in Fig.~\ref{eigen_sc-AA(AB)}(d). This manifests the enhanced broadening of zero-energy states in bilayer graphene SC, and it is consistent with the results of DOS in Fig.~\ref{DOS}(a). Therefore, the central peaks around the $E=0$ eV in Fig.~\ref{DOS}(a) correspond to these edge states, and such edge states are often caused by the zigzag edges of the honeycomb lattice, where atomic vacancies break lattice symmetry and induce the energy broadening. For SC-AB in Figs.~\ref{eigen_sc-AA(AB)}(e) and~\ref{eigen_sc-AA(AB)}(f), we can also observe similar distribution for high energy states. The energy broadening of the central peak also increases, as shown in the Figs.~\ref{eigen_sc-AA(AB)}(g) and~\ref{eigen_sc-AA(AB)}(h). Compared with SC-AA, the nonzero probability densities at Fermi energy are not only distributed at the internal lattice edge, but also located at the top and bottom boundary of the sample.

\begin{figure*}[tbp]
\centering
\includegraphics[width=1.0\textwidth]{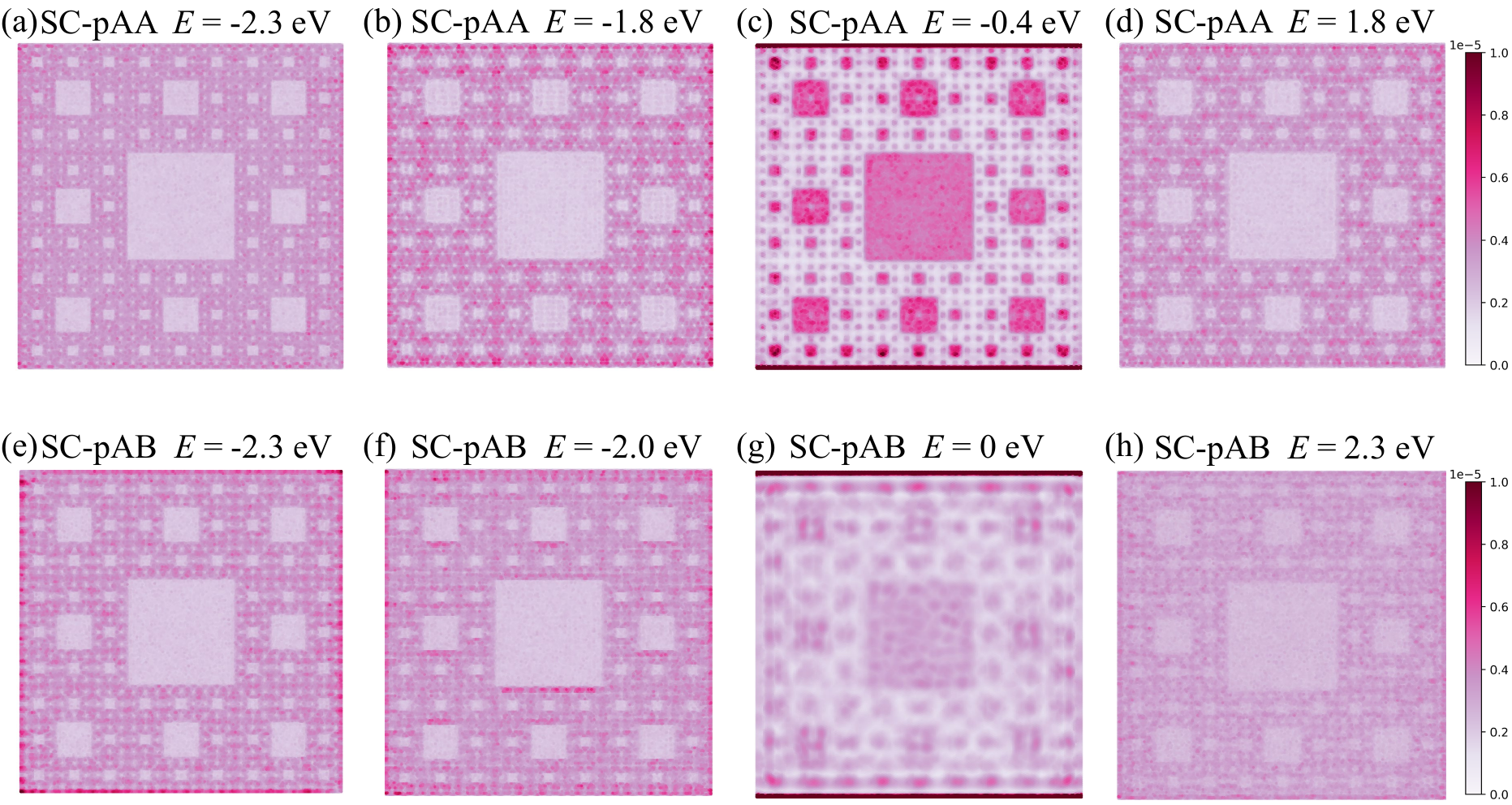}
\caption{The real-space distribution of quasi-eigenstates for SC-pAA under the pressure of 30 GPa at (a) $E=-2.3$ eV, (b) $E=-1.8$ eV, (c) $E=-0.4$ eV, and (d) $E=1.8$ eV. The real-space distribution of quasi-eigenstates for SC-pAB under the pressure of 30 GPa at (a) $E=-2.3$ eV, (b) $E=-2.0$ eV, (c) $E=0$ eV, and (d) $E=2.3$ eV. The sample parameters are set as $W=297.5$a, and $I=4$.} 
\label{eigen_sc-pAA(pAB)}
\end{figure*}

We next discuss the quasi-eigenstates of SC-pAA(pAB) under different pressures. First, we sum over the amplitudes of the normalized quasi-eigenstate in Area I without pressure, which can be a measure of the distributions in Area I. We call this quantity as the occupation percentage $O$. If $O$ of Area I for a given energy is $100\%$, it means that this given energy state is distributed only inside Area I. In this case, the electron with this energy is completely confined in the "fractal" region. On the contrary, if $O$ is 0, electrons are localized inside Area II, i.e., electrons have no access to any site in Area I. In Figs.~\ref{quasi_O}(a) and~\ref{quasi_O}(b), we show the calculated occupation percentage $O$ of quasi-eigenstates under different pressures in the energy range -2.5 eV to 2.5 eV. In SC-pAA, the $O$ maintains a high value and forms a platform with $|E|> 1.8$ eV, which means eigenstates at the high energy are mainly distributed in Area I. Starting from $|E|=1.8$ eV, $O$ begins to drop sharply, meaning that most of these quasi-eigenstates exist Area II. Interestingly, here the energy value $|E|=1.8$ eV corresponds exactly to the Van-Hoff singularities in the density of states of SC-pAA, which suggests that the energy corresponding to the Van-Hove singularity is the distribution transition interval where the region occupied by eigenstates begins to change. Besides, as the pressure increases, the value of $O$ becomes larger in the high energy region and smaller in the low energy region, resulting in a stronger localization of these states in corresponding energies. This manifests that greater pressure can make the electrons distributed in the fractal-like space, i.e., Area I. Near the Fermi level, we can see small peaks of $O$ due to the boundary states at the top and bottom of Area I in SC-pAA sample (see Fig.\ref{eigen_sc-pAA(pAB)}(c) below). In Fig.~\ref{quasi_O}(b), we can see similar behaviors of $O$ in SC-pAB. Electrons are mainly distributed in Area I inside the high-energy range. The energy around the Van-Hove singularity ($E=-2.0$ eV and $E=2.3$ eV) is the transition interval where the confined region of electrons will change from Area I to Area II or Area II to Area I. The peak around the Fermi energy $E=0$ eV represents the appearance of strong boundary states in SC-pAB (see Fig.~\ref{eigen_sc-pAA(pAB)}(g) below), which is different that in SC-pAA. We also plot the variation of $O$ of Area II under the different pressures for a fixed energy, as shown in Figs.~\ref{quasi_O}(c) and~\ref{quasi_O}(d). Taking the energy near the Van-Hove singularity as an example, the value of $O$ varies with pressure in an exponential curve distribution. It means that as the pressure increases the change of $O$ will gradually decrease.

For visualization, taking 30 GPa as an example, the real-space distributions of their quasi-eigenstates at several energy are shown in Fig.~\ref{eigen_sc-pAA(pAB)}. For high energy of SC-pAA in Figs.~\ref{eigen_sc-pAA(pAB)}(a), the electrons are mainly distributed inside Area I, suggesting that the electrons can be confined in fractal space. Even in the case of the transition energy, i.e. around the Van Hove singularity, the distribution remains in Area I, as shown in Figs.~\ref{eigen_sc-pAA(pAB)}(b) and ~\ref{eigen_sc-pAA(pAB)}(d). In contrast, for the low energy region, we can see that the eigenstate is localized in Area II in Fig.~\ref{eigen_sc-pAA(pAB)}(c), and the localized boundary states at the top and bottom of the sample are also manifested, which means that the electrons at the Fermi energy can not enter to the fractal space and are confined in the pressured region. Similar phenomenon also appears in SC-pAB, as shown in Figs.~\ref{eigen_sc-pAA(pAB)}(e) and~\ref{eigen_sc-pAA(pAB)}(f). At the Fermi energy, we can see that the localization of states in Area II and these boundary states at top and bottom side of sample, as shown in Fig.~\ref{eigen_sc-pAA(pAB)}(g). This also proves the central peak in Fig.~\ref{quasi_O}(b). For the transition energy point of $E=2.3$ eV, We can also see that the eigenstates are mainly distributed in Area I, and the proportion of distribution in Area II is not small. Therefore, We suggest that as the energy approaches the low-energy range, the distribution area of the eigenstates transfers from Area I to Area II.

Based on these results, we conclude that the distribution of quasi-eigenstates in SC-AA(AB) can confirm the energy broadening of central peaks in the energy spectrum increases. For the fractal-like pressure modulated bilayer graphene, although the distribution of quasi-eigenstates near the zero energy are mainly localized in Area II, high energy state can localizes fractal space (Area I). Within a certain range, stronger pressure can lead to stronger localization, forming a more efficient fractal space.

\subsection{Quantum transport}
In graphene SC systems, the geometry dimension can be revealed by the quantum conductance fluctuations by virtue of a box-counting (BC) method, i.e., the BC dimension of the quantum conductance fluctuations reflecting the Hausdorff dimension \cite{2016transport,yang2020confined,yang2022electronic}. However, for bilayer graphene fractals (SC-AA and SC-AB), the interlayer coupling may affect the quantum conductance fluctuations, and hence it is meaningful to discuss whether the correlation between the BC dimension and the Hausdorff dimension still exists.

\begin{figure}[H]
\centering
\includegraphics[width=8.4cm]{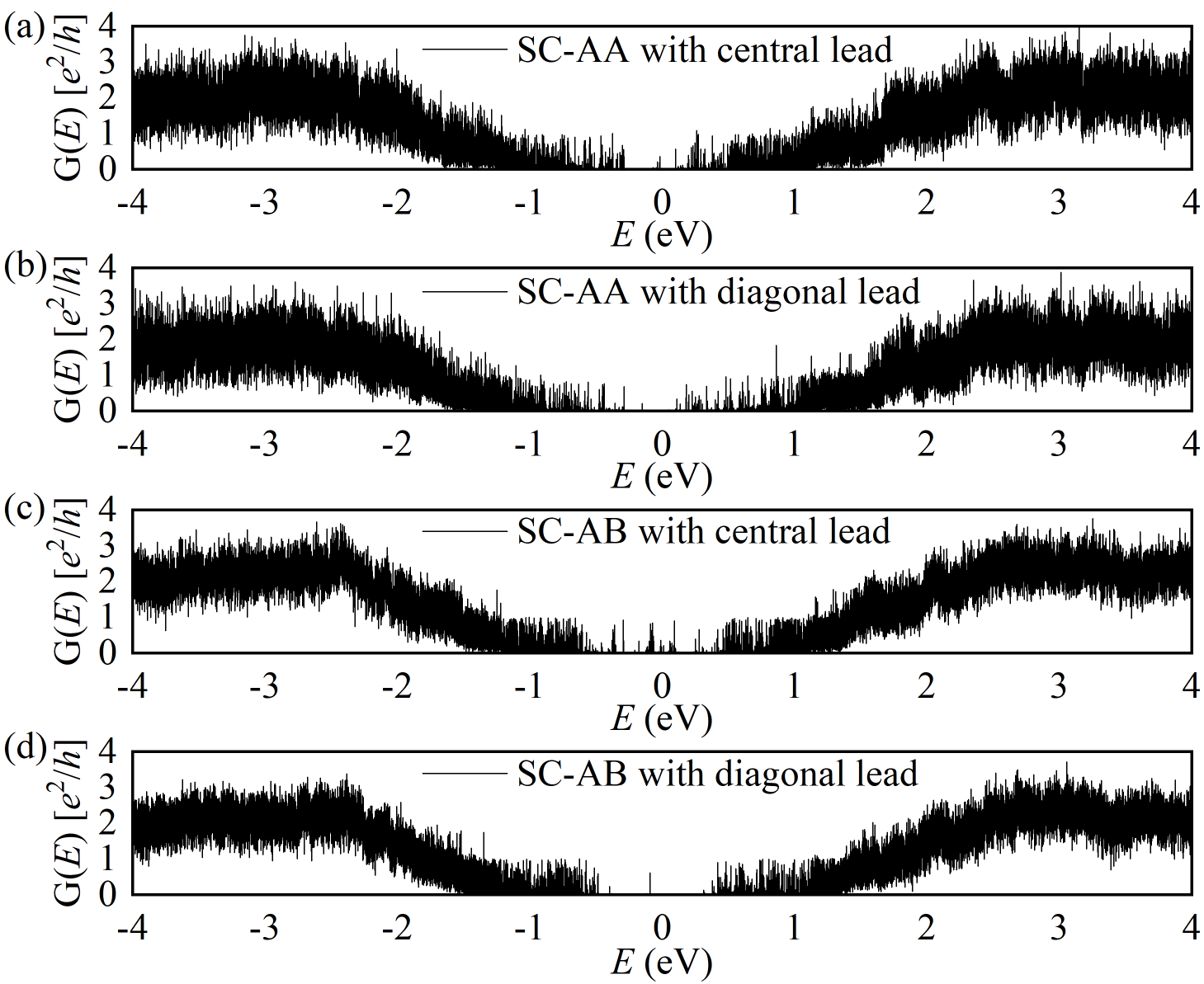}
    \caption{The conductance $G(E)$ (in units of $e^{2}/h$) of (a, b) SC-AA and (c, d) SC-AB as a function of energy under central/diagonal lead positions, where lead width is $(3+\frac{2\sqrt{3}}{3})a$. The sample parameters are set as $I=4$ and $W=297.5a$.}
    \label{conduc}
\end{figure}

The calculation of quantum conductance is implemented in the Kwant software by the Landauer formula of the scattering theory. The numerical value of the conductance can be changed by the number, position and width of the electrodes \cite{2016transport,yang2020confined}. We discuss the quantum conductance under two different lead position configurations, called the center leads (i.e., two leads are attached to the centrals of the left and right sides of sample) and the diagonal leads (i.e., one lead is attached to the bottom of one side and one lead is attached to the top of the other side), as shown in Fig.~\ref{sample}. The calculated quantum conductance spectra $G(E)$ in SC-AA and SC-AB with different lead configurations are shown in Fig.~\ref{conduc}. The conductance calculation of SC-pAA(pAB) are not performed, since the matrix dimension in SC-pAA(pAB) exceed the computational limits of Kwant. 

In monolayer graphene SC, there is a remarkable conductance gap in the central part of $G(E)$ where the conductance vanishes, which is a hallmark of electronic transports in SC fractals based on monolayer graphene \cite{2016transport,yang2020confined,yang2022electronic}. However, in SC-AA and SC-AB based on bilayer graphene, several minor conductance peaks exist inside low energy region in Figs.~\ref{conduc}(a)-~\ref{conduc}(c) because of the broadening of the central peaks near zero energy. The lead configurations also have remarkable different effects on the conductance inside low energy region. For the center lead case, the conductance gap almost vanish, while a narrow conductance gap exist especially for SC-AB in Fig.~\ref{conduc}(d) for the diagonal lead case. 

\begin{figure}[H]
\centering
\includegraphics[width=8.7cm]{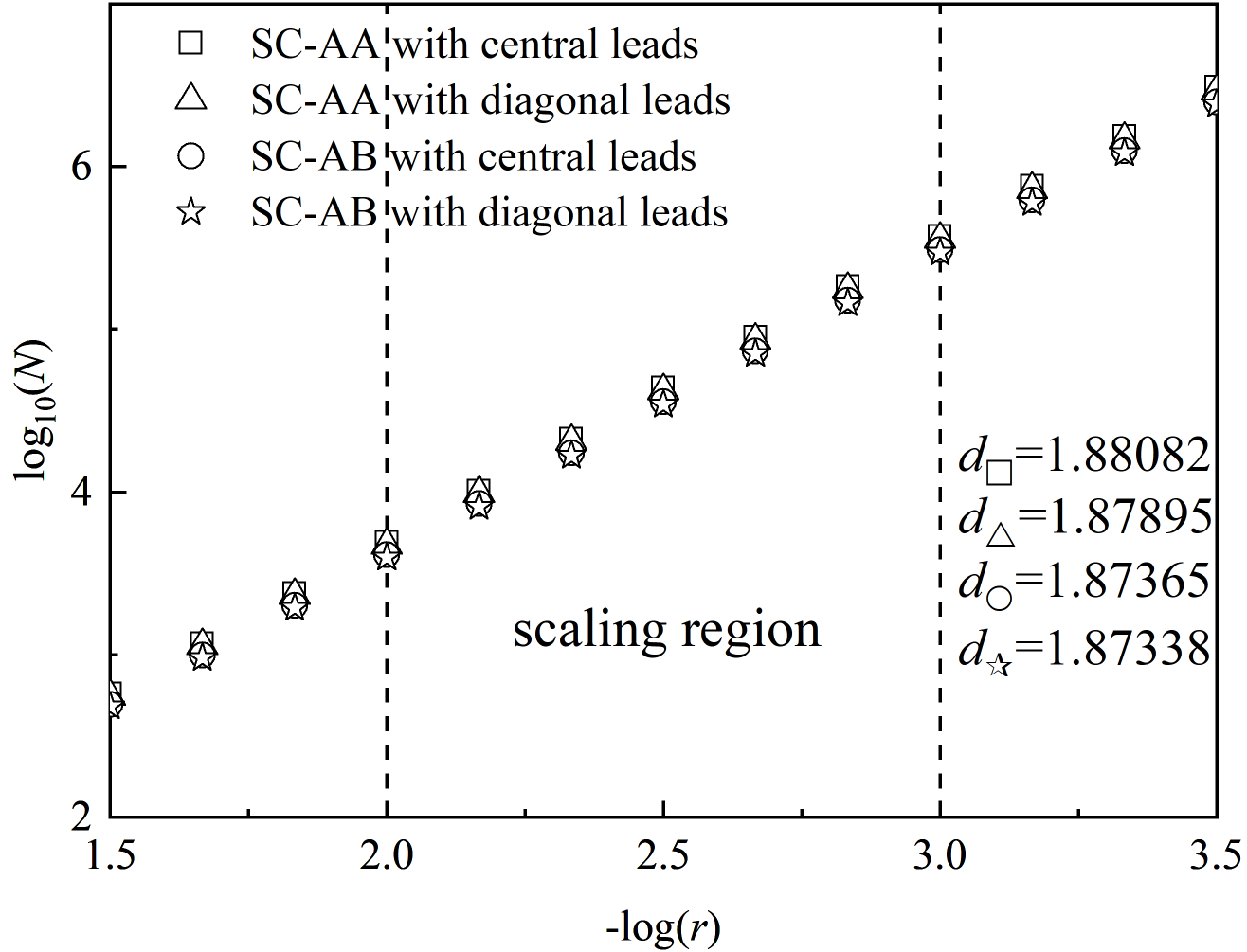}
    \caption{Box-counting algorithm analysis of the conductance fluctuations for SC-AA and SC-AB in Fig.~\ref{conduc}.}
    \label{BC_dimension}
\end{figure}

Beyond the low energy region, these conductance spectra contain many fluctuations making the curve quite noisy. The fluctuations can be characterized by the dimension of the whole conductance spectrum, which can be obtained by the box-counting algorithm \cite{guarneri2001fractal}. It is a pixel-based method and the core of this algorithm is to cover the data curve by boxes with a size $r$. The number of boxes $N$ depends on the size of $r$, and when $d=-\log_{10}(N)/\log_{10}(r)$ changes linearly, the value of $d$ is called as the BC dimension. In this situation, $d$ is the slope of the function $\log_{10}N[-\log_{10}(r)]$, and the region where $d$ change linearly is also called "scaling region". Actually, there are two other regions that have been dropped in the BC algorithm. For large values of $r$, where $-\log_{10}(r)$ is around 0, the box is too large to grasp the features of quantum conductance fluctuations, and for very small $r$, each box covers only one data point due to very small size so that $N$ is not increased anymore but turns to be a plateau. In Fig.~\ref{BC_dimension}, We show the numerical results of BC algorithm for SC-AA(AB). We consider the position effects of leads on the conductance spectrum and extract the BC dimension. For SC-AA(AB), the values of BC dimension are $d_{\square}=1.88082$ ($d_{\triangle}=1.87365$) and $d_{\bigcirc}=1.87895$ ($d_{\star}=1.87338$) in central and diagonal lead configurations, respectively. Surprisingly, the BC results in SC based on bilayer graphene are very close to the Hausdorff dimension $d_{\rm H}=1.89$. We can further infer that the slight difference between the box-counting dimension and Hausdorff dimension will vanish if the ramification number is infinite. This means that the correlation between quantum conductance and geometry dimension remains in spite of the existing interlayer coupling in bilayer graphene SC.

\section{SUMMARY}\label{sec:Summary}
We investigated the DOS and Quasi-eigenstates in SC-AA(AB) and SC-pAA(pAB) structures and the quantum conductance fluctuations in SC-AA(AB). The DOS results of SC-AA(AB) indicate the larger energy broadening of the edge states compared with monolayer graphene SC, and the DOS of SC-pAA(pAB) gradually become similar to SC-AA(AB) as pressure increases, but it can not exactly replicate the same spectrum of SC-AA(AB) within the experimental pressure range. The analyses on real-space distributions of normalized probability density for quasi-eigenstates also verified the DOS results. In SC-pAA, quasi-eigenstates around the zero energy are mainly localized inside Area II and the states inside high energy range are dispersed in fractal space (Area I). In SC-pAB, most quasi-eigenstates except these near Van-Hoff singularities are mainly localized inside fractal space. By summing over the amplitudes of the normalized quasi-eigenstates in Area I (i.e., occupation percentage), we find that within a certain pressure range, stronger pressure can lead to stronger localization, forming an efficient fractal space for high energy quasi-eigenstates. We Calculated the conductance spectrum in SC-AA(AB) and find that the quantum conductance fluctuations still follow the Hausdorff fractal dimension behavior. Thus, the high correlation between quantum conductance and geometry dimension is not affected in bilayer graphene SC in spite of the interlayer coupling.

\begin{acknowledgements}
This work was supported by the National Natural Science Foundation of China (Grants No. 12174291 and No. 12247101) and the Knowledge Innovation Program of Wuhan Science and Technology Bureau (Grant No. 2022013301015171). Y.W. acknowledges the support from the 111 Project (Grant No. B20063) and the National Key Research and Development Program of China (Grant No. 2022YFA1402704). We thank the Core Facility of Wuhan University for providing the computational resources.
\end{acknowledgements}

\bibliographystyle{apsrev4-2}
\bibliography{ref}

\end{document}